\documentclass[aps,prb,twocolumn,nobibnotes]{revtex4-1}
\usepackage{graphics,graphicx,amsfonts,amsmath,amsbsy,amssymb,color}
\usepackage{bm}
\usepackage[a4paper,vmargin={20mm,20mm},hmargin={20mm,10mm}]{geometry}
\usepackage{subfigure}
\usepackage{paralist}

\def \beq {\begin{eqnarray}}
\def \eeq {\end{eqnarray}}

\begin{document}
\title{Spectral functions of strongly correlated extended systems via an exact quantum embedding}
\author{George~H.~Booth}
\email{george.booth@kcl.ac.uk}
\affiliation{Department of Chemistry, Frick Laboratory, Princeton University, Princeton, New Jersey 08544, USA}
\affiliation{Department of Physics, King's College London, Strand, London WC2R 2LS, UK}
\author{Garnet~Kin-Lic~Chan}  
\affiliation{Department of Chemistry, Frick Laboratory, Princeton University, Princeton, New Jersey 08544, USA}

\begin{abstract}
Density matrix embedding theory (DMET) [Phys. Rev. Lett., {\bf 109}, 186404 (2012)], introduced a new approach to quantum cluster embedding methods,
whereby the mapping of strongly correlated bulk problems to an impurity with finite set of bath states was rigorously formulated to exactly reproduce the
entanglement of the ground state. The formalism provided similar physics to dynamical mean-field theory at a tiny fraction of the cost, but was inherently limited by 
the construction of a bath designed to reproduce ground state, static properties.
Here, we generalize the concept of quantum embedding to dynamic properties and demonstrate accurate bulk spectral functions at similarly small
computational cost. 
The proposed spectral DMET utilizes the Schmidt decomposition of a response vector, mapping the bulk dynamic correlation functions to that of
a quantum impurity cluster coupled to a set of frequency dependent bath states.
The resultant spectral functions are obtained on the real-frequency axis, without bath discretization error, and 
allows for the construction of arbitrary dynamic correlation functions.
We demonstrate the method on the 1D and 2D Hubbard model, where we obtain 
zero temperature, thermodynamic limit spectral functions, and show the trivial extension to two-particle Green functions. 
This advance therefore extends the scope and applicability of DMET in condensed matter problems as a computationally tractable 
route to correlated spectral functions of extended systems, and provides a competitive alternative to dynamical mean-field theory for dynamic quantities.
\end{abstract}
\date{\today}
\maketitle

\section{Introduction}

Dynamic correlation functions are directly probed in spectroscopic methods,
and describe the transport, optical, magnetic and wider electronic structure properties of materials. 
As such, their accurate computation is highly sought after. 
However, few robust approaches exist for strongly correlated materials\cite{Gali2013}. The difficulty is in simultaneously requiring both an accurate 
treatment of the electron correlations beyond mean-field density functional or low-order perturbation theory for the ground state {\em and} excitation 
spectrum, as well as modeling a system of sufficient size to reach the thermodynamic limit and avoid spurious finite size effects. 
In general, only mean-field electronic structure methods are computationally cheap enough to access the required system
sizes, and so there is a pressing need for methods with mean-field computational scaling, which can correctly describe 
the excitation spectrum in strongly correlated materials.

A zero-temperature dynamic correlation function can be defined in the frequency domain as
\begin{equation}
    G(\omega;{\hat X},{\hat V}) = \langle \Psi^{(0)} | {\hat X}^{\dagger} \frac{1}{\omega-(H-E_0)+i \eta} {\hat V} | \Psi^{(0)} \rangle , \label{eqn:intCorrFunc}
\end{equation}
with the one-particle Green functions defined with ${\hat V}$ and ${\hat X}$ being single annihilation/creation operators 
with appropriate time-ordering. 
Spectral functions are then defined as $A(\omega)=-\frac{1}{\pi}\Im[G(\omega)]$,
where the spectral broadening is given by the small imaginary component of the energy, $\eta$, which regularizes the correlation function. 
The single particle density of states, experimentally measured by scanning tunneling 
or photoemission spectroscopy, is the spectral function of the one-particle Green function. 
Two-particle spectral functions are also highly sought after, such as the two-hole propagator 
probed with Auger spectroscopy\cite{Mona2013}, or the particle-hole Green function that gives the optical conductivity, polarizability and
magnetizability, and are key descriptors in 
Raman spectroscopy and the mechanism of high-$T_c$ superconductivity\cite{Millis2012,Sordi2012,Millis2013}.

One method which has revolutionized the computational description of strongly correlated spectral functions is dynamical mean-field 
theory (DMFT, or Cluster-DMFT for a multi-site extension)\cite{Georges1992,Georges1996,Kotliar2006}. 
In DMFT, a single unit cell (or site) in the bulk is viewed as a quantum impurity, described by its local, one-particle Green function,
and is self-consistently embedded with a non-interacting self-energy (hybridization) from its environment.
The quantum impurity problem is then solved via an `impurity solver', such as
continuous-time quantum Monte Carlo (CT-QMC)\cite{Millis2006} or exact diagonalization (ED)\cite{Zgid2012}. 
A success of the approach is that the non-trivial physics in the limit of a high dimensional system can be solved exactly, as the correlated corrections to the
Green function are then local.
However, there are some formal drawbacks of the DMFT formulation. 
The essential problem is that DMFT maps the infinite bulk onto a quantum impurity plus infinite bath problem, which while obviously far
simpler, is still numerically challenging to solve.

If CT-QMC is used as an 
impurity solver to integrate over this infinite bath (or indeed general quantum Monte Carlo approaches in isolation), then spectral 
functions are obtained only on the imaginary frequency 
axis, requiring unstable analytic continuation onto the real frequency axis, which can wash out subtle or sharp features of the 
spectra\cite{Millis2006,Millis2009}. Accessing low temperatures or arbitrary interactions produces Fermion sign problems, which can require millions
of computer-hours to stabilize, even for a modest number of impurity sites.
On the other hand, exact diagonalization suffers from bath discretization error in the spectra due to the 
need for a second, approximate mapping of the hybridization (at all frequencies) to a representation by a 
finite number of bath sites 
with frequency independent energies and couplings\cite{Liebsch2012}. 
In addition, since DMFT is formulated via the Dyson equation on the one-particle Green function, other spectra such as the two-particle 
Green function and optical spectra are difficult to obtain, formally requiring expensive vertex corrections\cite{Millis2012}. 
Other methods to calculate 
spectra of correlated extended systems, such as the dynamical density matrix renormalization group\cite{Jeckelmann2004} or perturbative
methods\cite{Senechal2000} are often limited to certain correlation strengths, system sizes or spatial dimensions.

Here, we avoid many of these issues, by extending the idea of quantum embedding of {\em wavefunctions} rather than Green functions, to compute general dynamical
correlation functions. The framework, which we term density matrix embedding theory (DMET) was introduced for embedding of ground state wavefunctions in 
Ref.~\onlinecite{Chan2012}, with extensions to long-range Hamiltonians and broken symmetry ground states found in Refs.~\onlinecite{Chan2013,Chen2014,Bulik,Duchulkis}.
The essential idea for static quantities is that the Schmidt decomposition of a mean-field bulk state defines a manifestly {\em finite} quantum impurity mapping:
for a cluster of $l$ impurity sites there are (at most) $l$ bath sites. The mapping is exact in the non-interacting and local limits, similar to DMFT, but
has no bath discretization error. Furthermore, because it is a mapping to a finite quantum problem, it is orders of magnitude cheaper numerically to solve the
resultant quantum system, while the accuracy of DMET compares very favorably (and can exceed) that of DMFT.

Here, we show that for dynamic quantities, the Schmidt decomposition of a mean-field bulk response vector yields a similar finite quantum impurity mapping
that renders {\em spectra} exact in the non-interacting and local excitation limits. This `spectral DMET' possesses advantages compared to
DMFT. The finite quantum impurity mapping results in numerically simple calculations of spectral functions (no frequency point in these results
took more than a minute on a single computing core), while eliminating bath discretization error of DMFT+ED calculations.
In addition, since the method is formulated in terms of a general response vector, it is not restricted in the type or rank of perturbing   
operators that it can consider, with two-particle Green functions available within the same framework at a comparable computational 
cost to the single particle Green functions. A major contrasting feature of the approach compared to DMFT+ED is that the 
bath space and couplings change continuously with frequency, which is a major factor in the compactification of the bath. This frequency-dependent bath space
can then exactly represent the entanglement between the impurity and its environment at a given frequency, as defined by a global mean-field response function.
Furthermore, since this bath is not fit and can be algebraically constructed based on real frequency values, direct evaluation of the correlated spectral function
on the real frequency axis is straightforward.

The approach will be demonstrated for the Hubbard model, defined by the Hamiltonian in the site basis as
\begin{equation}
H = -t \sum_{\langle ij \rangle, \sigma} (a_{i,\sigma}^{\dagger}a_{j,\sigma} + \text{\textnormal {h.c.}}) + U \sum_i (n_{i,\uparrow} - \frac{1}{2})(n_{i,\downarrow} - \frac{1}{2})  . \label{eqn:hub}
\end{equation}
This is one of the key unresolved models in 
the electronic structure of correlated materials, encapsulating the competing effects of kinetic energy and Coulomb repulsion, which can give rise to correlation driven 
phase transitions and the Mott--Hubbard physics of many transition metal oxides\cite{Limelette2003}, including qualitative features 
of high-$T_c$ superconductivity\cite{Anderson87,Sordi2012,Millis2013}. Where possible, we compare to one-dimensional exact results from the Bethe 
Ansatz\cite{Lieb68,Ovchinni1970}, as well as benchmark 1D and 2D cluster DMFT calculations at half-filling 
and under doping\cite{Go2009,Kotliar2008,Masatoshi2009}. In addition, we compute local two-particle Green functions.

\section{Methodology}

We first recap the DMET quantum embedding formalism introduced for the ground state to familiarize the approach and enable similarities with the spectral adaptation to be emphasized, 
with more details provided in Ref.~\onlinecite{Chan2012,Chan2013} and \onlinecite{Duchulkis}. The 
quantum impurity model and bulk properties are defined in the following steps.
\begin{enumerate}
\item From the ground state of a one-particle Hamiltonian defined over the lattice, $h$, a single Slater determinant, $|\phi^{(0)}\rangle$, is variationally minimized.
\item A set of local `impurity' sites are chosen, spanned by the states $\{ |\alpha \rangle \}$. The formal Schmidt decomposition 
of $|\phi^{(0)}\rangle = \sum_{\alpha \beta} \phi_{\alpha \beta} |\alpha \rangle |\beta \rangle$ over this impurity space
yields a set of bath states, $\{ |\beta \rangle \}$, whose Fock space is the same dimension as the impurity space. For the case of the decomposition 
of a single Slater determinant, $\{ |\beta \rangle \}$ takes the form of a set of single particle states multiplied by a determinant of the states remaining 
in $|\psi^{(0)}\rangle$ which are rendered uncoupled via $h$ to the impurity space by the procedure.
\item The interacting quantum impurity plus bath Hamiltonian is constructed by projecting $H$ into this basis 
of $\{ | \alpha \rangle \} \otimes \{ | \beta \rangle \}$ as $H'=PH_{\mathrm emb}P$, 
with $P = \sum_{\alpha \beta} |\alpha \beta \rangle \langle \alpha \beta |$. This space spans the exact entanglement of the impurity to its environment 
within $|\phi^{(0)}\rangle$, and is now independent of the total number of sites in the system. To avoid double counting
of correlation effects, $H_{\mathrm emb}$ is defined as the fully-interacting $H$ over the impurity, while the one-particle $h'$ is used over the rest of the space.
\item $H'|\Psi^{(0)} \rangle = E_0 |\Psi^{(0)} \rangle$ is solved for the wavefunction over the quantum impurity and bath, $|\Psi^{(0)} \rangle$.
\item A one-particle, Hermitian potential defined over the impurity space, $u$, is obtained in order to match the elements of the one-body density 
matrix of $|\phi^{(0)}\rangle$ and $|\Psi^{(0)} \rangle$. This interaction potential mimics some of the longer-ranged correlation effects, and controls the
effective number of particles on the impurity cluster.
\item This defines the new one-particle lattice Hamiltonian, $h' = h + u$ (where $u$ is periodically repeated across the lattice), 
from which the process can be repeated until self-consistency.
\item Static, local expectation values are defined as $\langle \Psi^{(0)} | {\hat O} | \Psi^{(0)} \rangle$, while non-local expectation values, such as the 
energy for the impurity cluster, can be defined through a partial trace of the density matrices with the Hamiltonian, as defined in Ref.~\onlinecite{Chan2012,Duchulkis}.
\end{enumerate}

We now generalize to dynamic quantities. This requires the construction of a set of frequency-dependent bath states, into which the
interacting dynamic response equations are projected. This procedure exactly embeds the local spectrum in the response of the entire (formally infinite) lattice, 
as defined by a response vector, $|\phi^{(1)}(\omega) \rangle$ constructed from the one-particle Hamiltonian, $h'$.
\begin{enumerate}
\item The one-particle response vector is obtained from the solution to
\begin{align}
|\phi^{(1)}(\omega) \rangle &= \left[ \omega-(h^{\prime} -\varepsilon_0)+i\eta \right]^{-1} {\hat V} |\phi^{(0)}\rangle  \nonumber \\ 
                            &= {\hat R}(\omega) \sum_{\alpha \beta} \phi^{(0)}_{\alpha \beta} |\alpha \rangle |\beta \rangle    .
\end{align}
where ${\hat R}(\omega)$ is the response operator relating $| \phi^{(0)} \rangle$ and $| \phi^{(1)}(\omega) \rangle$. 
\item The {\em operator} $R(\omega)$ is decomposed into separate operators as $R(\omega) = \sum_i {\hat {\mathcal{A}}}^{(i)}(\omega) {\hat {\mathcal{B}}}^{(i)}(\omega)$, 
where ${\hat {\mathcal{A}}}^{(i)}(\omega)$ acts only on the local impurity states $\{ |\alpha \rangle \}$, and ${\hat {\mathcal{B}}}^{(i)}(\omega)$ on the states of $\{ |\beta\rangle \}$.
\item The Schmidt decomposition of $| \phi^{(1)}(\omega) \rangle$ takes the form
\begin{equation}
|\phi^{(1)}(\omega) \rangle = \sum_{\alpha \beta i} \phi^{(0)}_{\alpha \beta} {\hat {\mathcal{A}}}^{(i)} |\alpha \rangle {\hat {\mathcal{B}}}^{(i)} |\beta \rangle .
\end{equation}
$| \phi^{(1)}(\omega) \rangle$ is contained in the space $\mathcal{K}(\omega) = \{ |\alpha \rangle \} \otimes \{\hat {\mathcal{B}}^{(i)}(\omega) | \beta \rangle \}$ which 
then defines the projector $P(\omega) = |\mathcal{K}(\omega)\rangle \langle \mathcal{K}(\omega) |$, into which the interacting response equations are projected.
\item The interacting response vector, $|\Psi^{(1)} (\omega) \rangle$, is calculated from
\begin{equation}
    P \left[ \omega - (H_{\mathrm emb}-E_0) + i \eta \right] P | \Psi^{(1)}(\omega) \rangle = P {\hat V} P |\Psi^{(0)} \rangle   ,   \label{eqn:ExactResponse}
\end{equation}
which in this work is solved via an exact, iterative procedure\cite{Langou2005}.
In this work, $|\Psi^{(0)} \rangle$ was identical to that found from the ground-state DMET. However, $|\Psi^{(0)} \rangle$ can be reoptimized in the larger 
space of $\mathcal{K}(\omega)$ (which also fully spans the space of the original ground state wavefunction, included 
when ${\hat {\mathcal{B}}^{(i)}}={\bf 1}$). However, this was found not to 
qualitatively change the results, and so $|\Psi^{(0)}\rangle$ and $E_0$ are taken to be the ground-state wavefunction and energy in the 
original quantum impurity and bath space of $\{ | \alpha \rangle \} \otimes \{ | \beta \rangle \}$.
\item The final spectral function is obtained as $G(\omega; \hat{X}, \hat{V}) = \langle \Psi^{(0)} | P \hat{X}^{\dagger} P | \Psi^{(1)}(\omega) \rangle$.
\end{enumerate}

There are a number of points to note about the above construction. For simplicity, we restrict ourselves here to consider {\em local} spectral functions, where
${\hat V}$ and ${\hat X}$ are both closed operators within the space of $\{|\alpha \rangle \}$.
The set of operators ${\hat {\mathcal{B}}}^{(i)}(\omega)$ include the unit operator.
This ensures that
the space of the ground-state DMET is included within $\mathcal{K}(\omega)$. The rest of the ${\hat {\mathcal{B}}}^{(i)}(\omega)$ operators define frequency-dependent
bath states, which may not always be orthogonal, but nonetheless exactly `span' $|\phi^{(1)}(\omega) \rangle$ throughout the lattice. This ensures that the spectral 
functions are exact in the $U=0$ non-interacting limit, where $h'$, and therefore $|\phi^{(1)}(\omega) \rangle$ are also exact. Away from this limit, 
$|\phi^{(1)}(\omega) \rangle$ also changes, due to the presence of the interaction potential $u$ in the one-particle Hamiltonian, which 
includes local interaction effects of the extended system.
In addition, $|\Psi^{(1)} \rangle $ is also exact for uncoupled, local excitations within the impurity cluster, due
to the completeness of the impurity space $\{ |\alpha \rangle \}$ included in $\mathcal{K}(\omega)$.
One missing feature of the above construction is that the self-consistent potential $u$, is not updated as a function
of frequency, $u(\omega)$. This is appropriate at low energies, where the component of the ground state in the response function is large, but not at higher 
energies, as discussed further below. 

The exactness of the non-interacting and local limits is a feature that this construction shares with DMFT, but there are important properties which 
DMFT does not possess. The coupling to the environment
is achieved via a set of bath states, which change continuously with frequency, and which couple the impurity space to the set of 
non-local excitations given by the spectrum of $h^{\prime}$ for the frequency considered. 
In contrast, DMFT+ED constructs bath states which are
frequency-independent, and whose size must be formally infinite in order to avoid a discretization error which is not present in the above {\em algebraic} bath
construction. Furthermore, DMFT is formulated in terms of the one-particle Green function, but the procedure outlined above is suitable for 
general operators ${\hat V}$ of any rank. 
A key point of the approach is that the analytic construction
of the bath states which exactly span $|\phi^{(1)}(\omega) \rangle$, is no more costly than the diagonalization of the one-particle Hamiltonian. 
Once the fully interacting response is projected into this basis, there is no dependence on the size of the underlying lattice, rendering 
the cost of the method truly mean-field scaling with the size of the system.

\section{Results}

\begin{figure}
\begin{center}
\includegraphics[scale=0.435]{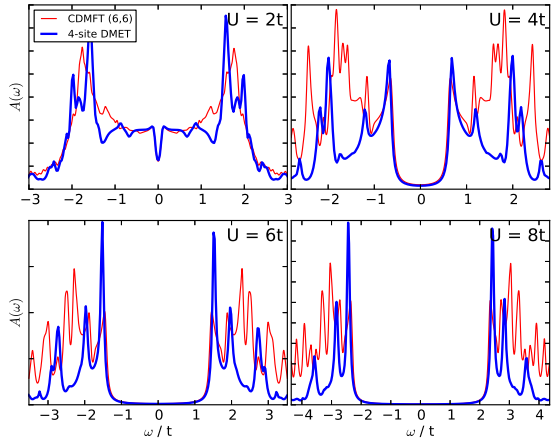}
\end{center}
\caption{Comparison of the local density of states between a four impurity cluster DMET calculation and a
(six impurity, six bath) CDMFT+ED calculation for the half-filled 1D Hubbard model. The analytic bath construction
of DMET renders the spectral functions smooth over the frequency range, with the same spectral broadening ($\eta=0.05t$) used for
both the CDMFT+ED and DMET calculations. }
\label{1D_DOS}
\end{figure}

We first examine the one-dimensional Hubbard model, whose ground state energy\cite{Lieb68} and spectral gap\cite{Ovchinni1970} are 
obtainable from the Bethe Ansatz, and additionally compare to
zero-temperature, cluster DMFT results\cite{Go2009}. Fig.~\ref{1D_DOS} shows the local density of states, calculated with a four-site DMET 
impurity cluster, and compared to six-site cluster DMFT results, obtained via an exact diagonalization within a six bath orbital representation. 
As expected in the one-dimensional case, there is no
frustration, and the system is dominated at all values of $U$ by long-range magnetic ordering\cite{Lieb68}. Consequently, 
the system is insulating for arbitrarily small values of $U$, in both the DMET and CDMFT spectra. 
The spectral gaps are very similar in the DMET and CDMFT calculations. However, many of the higher frequency peaks in the CDMFT calculations are very sharp -- these features
could be physical, or indeed could be spurious results of the finite (six) bath representation of the coupling of the cluster to the environment. In contrast, 
the DMET results give an entirely smooth
representation of the spectra, due to the analytic construction of the exact coupling to $|\phi^{(1)}(\omega)\rangle$ at each frequency. 

\begin{figure}
\begin{center}
\includegraphics[scale=0.435]{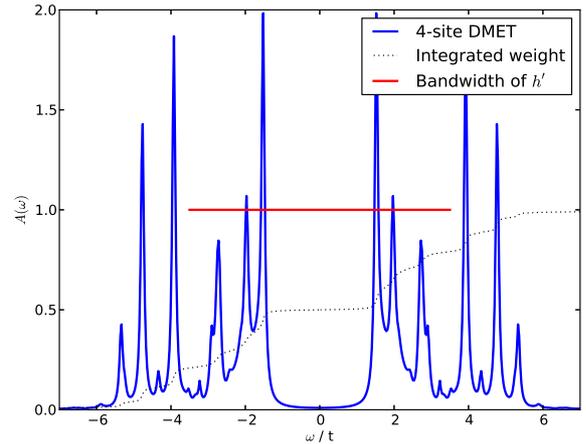}
\end{center}
\caption{Further detail of the $U=6t$ 4-impurity DMET spectra for the 1D Hubbard model. The dashed line gives the linetrace across all
    frequencies, demonstrating that the sum rule is obeyed, with the integrated weight appropriately unity over the entire frequency range. 
    However, at higher frequencies outside of the bandwidth of the mean-field response (shown in the plot), there is a lack of an appropriate 
    bath space to couple to high-energy local excitations, and high-weighted weakly coupled excitations can result, such as the ones between $4t\le|\omega|\le5t$.}
\label{1D_Peaks}
\end{figure}

\begin{figure}
\begin{center}
\includegraphics[scale=0.425]{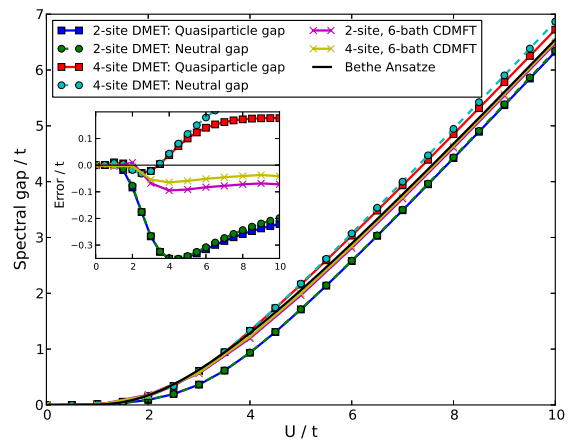}
\end{center}
\caption{Error in the spectral gap from the impurity 1-particle and 2-particle Green functions compared to analytic results
from the Bethe Ansatze\cite{Ovchinni1970}, and six bath orbital CDMFT+ED lattice Green functions. The results show generally 
good agreement, and convergence of the results with increasing cluster size. The spectral gap from the CDMFT calculations was taken
from the final lattice Green function, since this was found to be more accurate than the impurity Green function.}
\label{1D_GAP}
\end{figure}

However, at even higher frequencies, outside the bandwidth of $|\phi^{(1)}(\omega)\rangle$, there is little coupling of the local excitations to the 
rest of the lattice provided by the DMET bath construction. At larger values of $U$, this can result in uncoupled local excitations in the impurity sites, 
as shown explicitly in Fig.~\ref{1D_Peaks}. 
This is related to the frequency independence of the local potential $u$, and therefore inability of $h'$ to couple high energy excitations to the lattice
in a sufficiently physical manner.
While improving this aspect of the method will be a source of further work, in the following results we concentrate on the lower-frequency spectral window, 
determined by the bandwidth of $|\phi^{(1)}(\omega)\rangle$, ensuring an appropriate coupling of the excitations to the wider lattice response. 
Fig.~\ref{1D_GAP} shows the error in the spectral gap from the exact Bethe Ansatz result\cite{Ovchinni1970}, demonstrating 
the convergence towards the exact spectral gap as the cluster size is increased 
from two to four sites, and indicating a generally similar quality to CDMFT calculations.

\begin{figure}
\begin{center}
\includegraphics[scale=0.425]{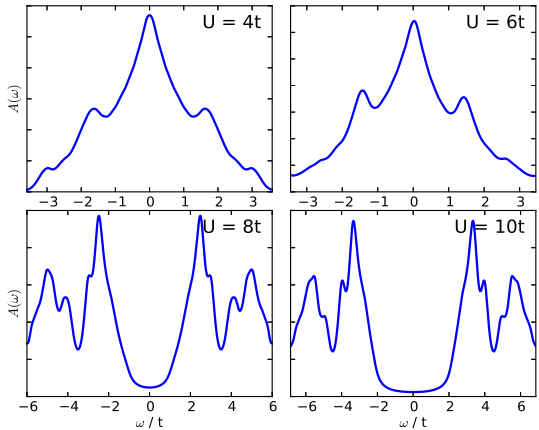}
\end{center}
\caption{Local density of states of the half-filled 2D Hubbard model from a $2 \times 2$ impurity cluster DMET calculation, with a 
MIT at $U\approx6.9t$. 
The exact bath states from $|\phi^{(1)}(\omega)\rangle$ for each frequency allows for smooth spectral functions at real frequencies.}
\label{2D_DOS}
\end{figure}

\begin{figure}
\begin{center}
\includegraphics[scale=0.425]{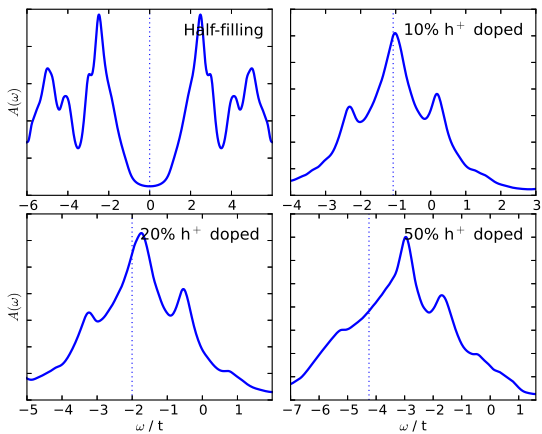}
\end{center}
\caption{Local density of states of the hole doped 2D Hubbard model calculated with a $2 \times 2$ impurity cluster with $U = 8t$. Dotted
lines indicate the Fermi level. Qualitative features present in the CDMFT study of Ref.~\onlinecite{Masatoshi2009} are observed (see Fig. 1(c) and (d)).}
\label{2D_Doped}
\end{figure}

A sterner test comes from the 2D paramagnetic Hubbard model. Cluster DMFT calculations exhibit a metal-insulator transition (MIT)
at finite $U$, representative of a Mott transition\cite{Georges1996,Kotliar2008,Millis2012,Millis2013}.
The DMET local density of
states are shown in Fig.~\ref{2D_DOS} for a $2 \times 2$ plaquette of impurity sites. In the low $U$ regime, the famous `three-peak' structure 
is observed, with a central Kondo resonance peak, and the Hubbard bands either side. At $U \approx 6.9t$, there is a transition to an insulating 
regime, with a Mott gap opening with increased $U$. The spectra then feature prominent coherence peaks at the gap edge\cite{Kotliar2008}.
Cluster DMFT calculations with the same plaquette size observe a MIT at a slightly lower $U\approx5.5t$, 
with a coexistence region at low temperatures, which we also observe. However, analytically continued
spectral functions from CT-QMC smooth out many of the subtle correlation driven substructures observed in Fig.~\ref{2D_DOS}. 
Fig.~\ref{2D_Doped} shows the effect on the 
spectra upon hole doping the system, which is shown for the insulating phase at $U=8t$. Once doped, the weight is transferred to lower frequencies
until the system becomes a Fermi liquid phase\cite{Masatoshi2009}. 

\begin{figure}
\begin{center}
\includegraphics[scale=0.425]{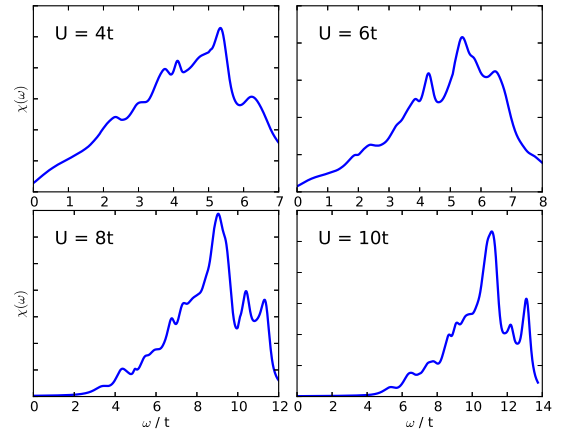}
\end{center}
\caption{$2 \times 2$ impurity DMET calculation of the local density-density response function for the half-filled 2D Hubbard model 
(${\hat V}={\hat X}=\sum_{\sigma} a_{\alpha_{\sigma}}^{\dagger}a_{\alpha_{\sigma}}$).}
\label{2D_DD}
\end{figure}

Fig.~\ref{2D_DD} shows the local density-density response function of the 2D Hubbard model at half-filling, obtained from the local two-particle Green 
function. The poles of this 
function correspond to the neutral excitation energies, and as such determine 
the optical properties of the system\cite{Millis2012,Essler91}.
Unfortunately, direct comparison of the spectra in Fig.~\ref{2D_DD} to other results is difficult due to the paucity of other comparable 
calculations for this quantity, but show the expected MIT, and transfer of spectral weight to higher frequencies as $U$ increases.
However, in the 1D Hubbard Hamiltonian
the optical and single-particle gaps obtained should be the same, and these can be tested against the exact Bethe Ansatz results
of Fig.~\ref{1D_GAP},
and show that the differences between the spectral gaps are generally small\cite{Essler91}.
While not conclusive, this supports the assertion that the one- and two-particle Green functions are of similar quality, 
while also of comparable computational cost, and validating the results of Fig.~\ref{2D_DD}.

\section{Conclusions}

In this manuscript, we have presented a quantum embedding formalism for accurate, zero-temperature spectral functions of extended, strongly correlated systems 
at a small computational cost, which vastly 
extends the scope of the DMET method. The approach has a number of 
advantageous formal properties, in particular when compared to DMFT-like theories. 
\begin{inparaenum}[\itshape i\upshape)]
\item The embedding within the environment is achieved via coupling to a finite set of analytically constructed, many-particle bath states,
    which are derived from the formal Schmidt decomposition of the response vector obtained from a one-particle Hamiltonian.
    The quantum impurity plus bath space therefore exactly spans this vector, and
    changes with frequency, to give smooth spectra without any error from discretization of the continuum. This contrasts with the formally infinite bath of DMFT.
\item The simple finite quantum impurity model allows spectral functions to be easily obtained on the real frequency axis, removing the 
    need for any analytic continuation from the imaginary axis.
\item The interacting quantum impurity and bath space is independent of the size of the underlying lattice, and is constructed at a cost no greater than the
    diagonalization of the one-particle Hamiltonian. The overall computational cost of this method is therefore very small compared to equivalent DMFT calculations.
\item Both extension to clusters of impurity sites, and arbitrary dynamic correlation functions are straightforward within the framework.
\end{inparaenum}

\acknowledgements

The authors would like to sincerely thank Ara Go for sharing CDMFT results and useful 
conversations, and Sebastian Wouters, Cedric Weber and Qiaoni Chen for useful comments on the manuscript.
This work was supported by the DoE through grants DE-SC0010530, and DE-FG02-12ER46875.

%

\end{document}